%% file: main_revised.tex
\documentclass[onecolumn]{aastex631}

\usepackage{comment}

\pdfoutput=1

\newcommand{\dgg}{$^{\circ}$}
\newcommand{\dg}{$^{\circ}$ }

\received{August 24, 2023}
\revised{September 29, 2023}
\accepted{October 16, 2023}

\shorttitle{ICME Magnetic Topology and GCR Modulation}
\shortauthors{E.E. Davies et al.}
\graphicspath{{./}{figures/}}
\begin{document}

\title{The effect of magnetic field line topology on ICME-related GCR modulation}

% Characterising Fds in MESSENGER data (at Mercury) associated with ICMEs
% ICME related Fds at Mercury from MESSENGER: one or two-step structure? 

\author[0000-0001-9992-8471]{Emma E. Davies}
\affiliation{Institute for the Study of Earth, Ocean, and Space, University of New Hampshire, Durham, New Hampshire, USA}
\affiliation{Austrian Space Weather Office, GeoSphere Austria, Graz, Austria}
\correspondingauthor{Emma E. Davies}
\email{emma.davies@geosphere.at}

\author[0000-0002-5681-0526]{Camilla Scolini}
\affiliation{Institute for the Study of Earth, Ocean, and Space, University of New Hampshire, Durham, New Hampshire, USA}

\author[0000-0002-9276-9487]{Réka M. Winslow}
\affiliation{Institute for the Study of Earth, Ocean, and Space, University of New Hampshire, Durham, New Hampshire, USA}

\author[0000-0002-0757-6382]{Andrew P. Jordan}
\affiliation{Institute for the Study of Earth, Ocean, and Space, University of New Hampshire, Durham, New Hampshire, USA}

\author[0000-0001-6868-4152]{Christian Möstl}
\affiliation{Austrian Space Weather Office, GeoSphere Austria, Graz, Austria}

%%%%%%%%%%%%%%%%%%%%%%%%%%%%%%%%%%%%%%%%%%
\begin{abstract}

The large-scale magnetic structure of interplanetary coronal mass ejections (ICMEs) has been shown to affect the galactic cosmic ray (GCR) flux measured in situ by spacecraft, causing temporary decreases known as Forbush decreases (Fds). In some ICMEs, the magnetic ejecta exhibits a magnetic flux rope (FR) structure; the strong magnetic field strength and closed field line geometry of such ICME FRs has been proposed to act as a shield to GCR transport. In this study, we identify four ICMEs near Earth that drove Fds with similar mean magnetic field strengths (20 -- 25~nT); two ICMEs with more typical mean speeds ($\sim$400~km~s$^{-1}$), and two fast ($\sim$750~km~s$^{-1}$) ICMEs. Within each speed pairing, we identify an ICME that exhibited an open magnetic field line topology and compare its effect on the GCR flux to that which exhibited a mostly closed topology. We investigate the different mechanisms that contribute to the resulting ICME-related Fds and their recovery, and determine which properties, if any, play a more important role than others in driving Fds. We find that much of the GCR response to the ICME events in this study is independent of the open or closed magnetic field line topology of the flux rope, and that features such as the fluctuations in speed, magnetic field structure, and expansion within the FR may play more of a role in determining the smaller-scale structure of the Fd profile.

\end{abstract}


\section{Introduction} \label{sec:intro}

Forbush decreases \citep[Fds;][]{forbush1937effects} are temporary decreases in the flux of galactic cosmic rays (GCRs) measured in situ by spacecraft.
Fds are associated with the arrival of different solar wind structures, specifically interplanetary coronal mass ejections (ICMEs) -- corresponding to the interplanetary counterparts of CMEs observed in the solar corona \citep{webb2012cmes}, and stream interaction regions (SIRs) generated by the interaction between fast and slow solar wind streams. Generally, the passage of such transients causes a rapid decrease in the GCR flux, followed by a slower recovery phase typically lasting many days \citep[e.g.][]{cane2000coronal}. In this study, our primary interest is the modulation of GCRs induced by the large-scale magnetic structure of ICMEs.

ICMEs present a magnetically-dominated core structure characterized by a high magnetic field, known as the magnetic ejecta (ME). MEs may exhibit signatures consistent with a magnetic flux rope (FR) structure \citep[a helical magnetic field wrapping around a central axis;][]{klein1982interplanetary, kilpua2017coronal}, and the common presence of counter-streaming electron strahls, known as bi-directional electrons (BDEs), suggest the large-scale magnetic structure of MEs is connected to the solar corona through two ``legs" \citep[see, e.g., Figure 2 in][]{zurbuchen2006situ}. Such MEs, with a smoothly rotating enhanced magnetic FR in addition to low plasma $\beta$ and low proton temperatures, are known as magnetic clouds (MCs) \citep{burlaga1981magnetic}. Additionally, ICMEs that are faster than the local magnetosonic speed in the solar wind reference frame drive forward shocks and sheaths \citep{richardson2010near, kilpua2017coronal, jian2018stereo}.

As ICMEs propagate, they undergo many processes including interactions and magnetic reconnection with the interplanetary magnetic field (IMF) in large-scale solar wind structures and other solar transients \citep[e.g.][]{winslow2016longitudinal,manchester2017physical, davies2020radial, davies2021solo, winslow2022, Scolini2022} . 
While BDE signatures within an ME can be interpreted as evidence of a magnetic field topology fully connected to the Sun, uni-directional electron signatures are traditionally interpreted as an indication of open field lines within an ME. Furthermore, heat flux dropouts (i.e., a complete lack of suprathermal electrons) can indicate complete disconnection of the field lines from the Sun. 
Uni-directional suprathermal electrons and flux dropouts in an ME can be yielded by various types of magnetic reconnection events. For example, \citet{dasso2006new} and \citet{ruffenach2012multispacecraft} distinguished between magnetic erosion (i.e., interplanetary reconnection) and interchange reconnection (occurring below the Alfv\'en point in the solar corona) largely based on the location of uni-directional suprathermal electrons (i.e., whether at the center or the front/back of an ME). More recently, \citet{winslow2023} proposed uni-directional electrons flowing in a constant direction (i.e., not alternating direction) and a lack of in situ reconnection signatures as additional requirements for distinguishing between erosion and interchange reconnection.

The modulation of GCRs by ICMEs, and therefore the resulting Fd profile, has been linked to the different substructures comprising an ICME. Fds frequently display two different decreasing trends between onset and minimum, either characterized by different gradients, or separated by a leveling of the GCR flux \citep[e.g.][]{cane1994cosmic}. Such Fd profiles are known as two-step Fds, where the initial decrease in GCR flux is often associated with the sheath region of the ICME \citep[see discussion in][]{vonforstner2020comparing}, and the second step associated with the magnetic structure of the ICME ejecta \citep[e.g.][]{lockwood1991forbush, cane1993cosmic}. This picture suggests that if only one ICME substructure was observed, a one-step Fd structure with a steady decrease in GCR flux from onset to the minimum of the Fd profile would occur. The frequency of one or two-step Fds, and thus which is most commonly observed, is a subject of debate with studies suggesting that the two-step structure of a Fd is not always clearly observed and may be less commonplace than previously suggested \citep{jordan2011revisiting}. To better understand the role different substructures play in driving the Fd, studies have performed superposed epoch analyses (SEAs) to characterize the average profile of Fds both near Earth using neutron monitor and ACE data \citep{masias2016superposed, janvier2021twostep}, and Mercury using MESSENGER neutron spectrometer data \citep{davies2023messenger}. These studies showed a clear link between the ICME magnetic boundaries of the shock, sheath, and magnetic ejecta regions to the different steps of the Fd profiles observed.

The mechanisms that affect the modulation of GCRs can be differentiated for the different ICME substructures and physical properties of each region \citep[e.g.][]{barnden1973largescale, barnden1973forbush}. The first feature to affect the GCR flux is the ICME shock, which acts as a discontinuity that can reflect particles causing an initial increase in the GCR flux observed just prior to the arrival of some ICMEs and contribute to the initial decrease in GCR flux after the ICME arrival \citep{kirin2020interaction, cane2000coronal}. Next, the turbulence of the sheath region acts as a diffusive barrier and affects GCR transport across the region associated with the first decrease in GCR flux \citep[e.g.]{wibberenz1997twostep, wibberenz1998transient}; thus models often focus on solving the diffusion-convection equation \citep[e.g.][]{bland1976simple, nishida1982numerical}. The subsequent decrease is associated with the strong magnetic field strength and closed field line geometry of the ME FR which have been proposed to act as a shield to GCR transport \citep{krittinatham2009drift}. Models of such GCR transport mechanisms consider the perpendicular diffusion of particles that fill the closed magnetic structure of the FR, solving the diffusion equation \citep[e.g.][]{cane1995response, richardson2011galactic, dumbovic2018analytical, dumbovic2020evolution}. Finally, after the passage of the ME, the recovery phase of the Fd may still be affected by the evolution of the disturbance caused by the ICME with heliocentric distance, effectively creating a ``shadow effect'', i.e., the recovery of the GCR flux depends on the exponential decay of the disturbance as it propagates away and particles propagate back into the shadow cast by the ICME \citep[see discussion in][]{lockwood1986characteristic, dumbovic2011cosmic}.

Models of perpendicular diffusion across the ME suggest the maximum decrease in GCR flux of the Fd can be described as proportional to $Bva^2$ \citep{cane1995response}, where $B$ is the magnetic field strength of the ME, $v$ is the ICME speed, and $a$ is the radius of the flux rope. These properties have also been found to affect Fd size in observational and statistical studies of ICME-related Fds. For example, \citet{winslow2018window} found that the magnitude of the ICME magnetic field played a significant role in modulating GCRs when investigating the evolution of an ICME-related Fd from Mercury, to Earth, and then to Mars. Furthermore, \citet{janvier2021twostep} found that by comparing ICMEs with and without sheath regions of similar speed profiles, the intensity of the magnetic field strength is the necessary condition in driving a Fd. In a similar study, \citet{masias2016superposed} investigated how the velocity of the ICMEs affected the average profile of the Fd, finding that the percentage decrease in GCR flux and the recovery period were larger for faster ICMEs. 

It has been hypothesized that the alteration of the ICME magnetic topology due to magnetic reconnection i.e. the opening of the closed magnetic field configuration of the ME flux rope, may also have a strong effect on the ICME’s ability to modulate GCRs \citep[e.g.][]{richardson2011galactic}. Based on models of GCR transport within the FR, one may expect open field lines to fill up more rapidly with GCRs than closed field lines which act as a shield to GCRs, and therefore, expect a lesser overall decrease in GCR flux within the ME and even brief periods of recovery in the GCR flux during periods of open field lines within the FR \citep{bothmer1997effects}. To investigate such an effect, \citet{richardson2011galactic} performed a statistical study of more than 340 ICMEs and their properties, in this case the percentage of bi-directional electrons observed within the ME, and the resulting Fd properties observed near/at Earth. They found only a slight trend towards more magnetically closed ICMEs being associated with greater GCR flux decreases. This trend was more evident for ICMEs with bi-directional electrons observed for $>90$\% of the ME duration. However, \citet{richardson2011galactic} also found a strong correlation between the fraction of closed field lines and the measured ICME speed, thought to be due to there being less time for reconnection of magnetic field lines to take place for faster ICMEs. Based on these findings, it is therefore difficult to discern whether the percentage of closed field lines directly controls the Fd size observed, or whether the speed of the ICME plays a more dominant role.   

To investigate this result further, in this study, we identify four ICMEs with strong mean magnetic field strengths ($|B_{mean}|>$ 20~nT) observed at L1 that drove Fds, as observed by neutron monitors at Earth. We first choose pairs of events with similar ICME speeds e.g. two ICMEs with more typical mean speeds, and two ICMEs that would be considered very fast ICMEs. We note that although the mean magnetic field strengths are similar across all events, the fast ICMEs have a much stronger maximum magnetic field strength ($|B_{max}|>$ 40~nT) than the more typical speed ICMEs ($|B_{max}|> \sim$~25~nT). Within each speed pairing, we identify an ICME that exhibited an open magnetic field line topology (i.e., one that likely underwent reconnection) and we compare its effect on GCRs to the other that exhibited a mostly closed topology (both ends likely connected to the Sun). By focusing on individual events, we are able to investigate the different mechanisms that contribute to the resulting ICME-related Fds and their recovery, and determine which properties, if any, play a larger role than others in driving Fds. 

This paper is organized as follows. In Section~\ref{sec:methods} we introduce the datasets used for our analysis, and the methods employed to prepare the data and analyze the selected ICME events. In Section~\ref{sec:identification}, we describe the criteria for the selection of appropriate ICME events, and we proceed with a detailed description of the selected events in Section~\ref{sec:description}. In Section~\ref{sec:discussion}, we compare the events and mechanisms of the GCR response, and finally, we draw conclusions summarizing our findings in Section~\ref{sec:conclusion}.

\section{Data and Methods} \label{sec:methods}

\subsection{Data used}
\label{subsec:data}

To identify the ICMEs in situ near Earth, we primarily use measurements of the magnetic field taken by the magnetometers onboard ACE \citep[Magnetic Field Experiment, MAG;][]{smith1998ace} and measurements of the solar wind plasma properties and suprathermal electrons \citep[Solar Wind Electron Proton Alpha Monitor, SWEPAM;][]{mccomas1998solar}. If ACE data is unavailable, we instead use measurements of the magnetic field taken by Wind \citep[Magnetic Field Investigation, MFI;][]{lepping1995wind}, complemented by measurements of the solar wind plasma \citep[Solar Wind Experiment, SWE;][]{ogilvie1995swe} and suprathermal electrons \citep[Three-Dimensional Plasma and Energetic Particle Investigation, 3DP;][]{lin1995wind3dp}. 

The corresponding Fds observed at the Earth are identified using measurements of the GCR count rate (counts per second) obtained by the South Pole (hereafter SOPO) %and McMurdo (hereafter MCMU) 
Neutron Monitor station as part of the Bartol neutron monitor project. SOPO is located at a latitude of $-90$\dg and altitude of 2820 m, with a cut-off rigidity of 0.10 GV. The cut-off rigidities, or atmospheric cutoffs, are the lower energy limit of GCRs that are able to reach the neutron monitors on the ground, and thus such cut-off rigidities decrease with altitude so as high-altitude polar neutron monitors are more sensitive to lower-energy particles \citep[see][for more details]{poluianov2022cosmic}. In conditions without a significant solar energetic particle event (GCRs only), \citet{poluianov2022cosmic} estimated the atmospheric cutoff energy at SOPO to be 322.3~MeV. In this study, we use 10 minute resolution pressure- and efficiency-corrected data obtained from the Neutron Monitor Database (NMDB) Event Search Tool (NEST; \url{https://www.nmdb.eu/nest/}).

\subsection{Time-shifting}
\label{subsec:time_shifting}

As Wind and ACE are located upstream of the Earth's magnetosphere at L1, it is necessary to time-shift the magnetic, solar wind plasma, and suprathermal electron data to directly compare features with those observed at Earth in the GCR data. To time-shift data from L1, we need to consider the separation between the Earth and the spacecraft, the Earth's orbital motion, and the speed and geometry of the solar wind phase front to be propagated. Assuming the solar wind phase front is planar, normal to the ecliptic, and propagates radially with the solar wind speed, we input the solar wind velocity and geometric considerations at each timestamp to the same time delay equation used to time-shift OMNI data using spacecraft located at L1 to the Earth bow shock (see \url{https://omniweb.gsfc.nasa.gov/html/omni_min_data.html} for details): 

\[\Delta_t(t) = {{X(t) + Y(t)W(t)} \over {v(t) - v_e W(t)}}, \]

\noindent where $\Delta_t$ is the resulting time-shift, $X$ and $Y$ are the position components of the spacecraft in Geocentric Solar Ecliptic (GSE) coordinates, $v$ is the observed solar wind speed, $v_e$ is the speed of Earth's orbital motion set to 30 km~s$^{-1}$, and $W$ is the assumed orientation of the phase front relative to the Earth-Sun line, $W = tan[0.5 tan^{-1}(v(t)/428 km~s^{-1})]$.

To calculate the time-shift to apply to each datapoint within the ICME, first we create a pseudo-velocity profile by linearly fitting the proton velocity throughout the ME. To extend the profile to the sheath region, we use the mean sheath speed calculated between the shock front and the start of the ME for each timestamp. However, if the maximum value of the linear fitting at the front of the ME is greater than the mean velocity of the sheath, we apply this value across the sheath region instead to avoid the possibility of datapoints with higher speeds `overtaking' preceding datapoints. By creating such a profile, we are able to capture the expansion of the ICME as it propagates toward the Earth.

The magnetic field, solar wind plasma, and suprathermal electron data were resampled to a 2 minute resolution, and the time-shift equation applied to each datapoint using the corresponding value of the reconstructed velocity profile, $V$. The resulting time-delta was then applied to each datapoint throughout the ICME. We also linearly time-shifted the data outside of the ICME by applying the same time-delta calculated at the shock front to the preceding data, and the time-delta calculated at the trailing edge of the ME to the following data. 

\subsection{Characterization of suprathermal electron signatures}
\label{subsec:BDE_identification}

To investigate the magnetic connectivity of ICMEs back to the Sun and its influence on the modulation of GCRs induced by individual ICMEs, we make use of ACE SWEPAM suprathermal electron pitch angle distributions (PADs) from the ACE Science Center (from \url{https://izw1.caltech.edu/ACE/ASC/DATA/level3/swepam/data/}), which are provided at the energy channel centered at 272~eV. For Wind, we use 3DP EHPD suprathermal electron PADs between 136~eV and 2~keV (from \url{http://sprg.ssl.berkeley.edu/wind3dp/data/wi/3dp/ehpd/}). In this case, we integrate the electron energy--angle distributions across different energy channels to obtain a single PAD map (particle flux distribution in time and angular direction) for all energy channels. 
The choice of Wind or ACE PAD data is based on resolution and data quality. ACE provides PAD information at 64 second cadence and across 20 angular channels each $9^\circ$ wide, that is, at higher temporal and angular resolutions than Wind with a 100 second cadence across 8 angular channels each $22.5^\circ$ wide. However, ACE PAD data are significantly degraded for events after 2011, so in these cases, we prefer Wind data.

For each event under study, we identify periods of bi-directional electron (BDE) signatures within the corresponding FR as follows.
At each time step for which data is available, we calculate the mean particle fluxes observed in the directions parallel ($I_{0}$, defined to be between $0^\circ$ and $45^\circ$), anti-parallel ($I_{180}$, between $135^\circ$ and $180^\circ$), and perpendicular ($I_{90}$, between $63^\circ$ and $117^\circ$ for ACE, and $\sim 67.5^\circ$ and $\sim 112.5^\circ$ for Wind) to the local magnetic field direction. The particle distribution at a given time step is marked as bi-directional if the ratio of the mean flux intensities in the parallel and anti-parallel directions is smaller than or equal to 2 (that is, the two strahls have similar intensities), and the ratio of the mean flux intensities from the weakest strahl direction to the perpendicular direction is larger than or equal to 2 (ensuring that the flux in both strahls is larger than the flux coming from the direction perpendicular to the local magnetic field). Different thresholds for the above ratios have been also tested, and the final thresholds have been selected as the ones proving most reliable for the identification of BDEs based on a comparison with a visual identification. Similar BDE identification criteria were also employed by \citet{richardson2011galactic}.

As a relevant metric for the selection of the events (Section~\ref{sec:identification}), we compute the percentage of the FR that exhibits BDE signatures in order to distinguish between ICMEs that exhibit predominantly open magnetic topologies (i.e., ones that likely underwent reconnection) and ICMEs that exhibit primarily closed magnetic topologies (both ends connected to the Sun). By doing so, we will test whether a connection/disconnection of ICME magnetic structures from the Sun actually associates with a greater/lesser ability of ICMEs to modulate GCR flux, or whether the magnetic connectivity is not a primary factor in controlling the resulting Fds.

\subsection{ICME interaction history}
\label{subsec:interaction_history}

Characterizing the interactions underwent by individual ICMEs during their propagation from the Sun to the Earth is important for the interpretation of the GCR and PAD signatures observed in situ at 1~au, and to clarify the role of interactions (e.g., through magnetic reconnection) in controlling the ICME's magnetic connectivity and GCR modulation. In this work, we search for possible interactions of each ICME with other solar wind structures, namely high speed streams (HSSs), stream interaction regions (SIRs), other ICMEs, the heliospheric current sheet (HCS)/ heliospheric plasma sheet (HPS), as well as isolated interplanetary shocks, that might have occurred during the ICME propagation from Sun to Earth. To do so, we investigate the presence of solar wind structures interacting with each ICME based on in situ plasma and magnetic field data available at L1. For the identification of interplanetary shocks, we cross-check our identifications with the Heliospheric Shock Database (hereafter ``IPShocks''), generated and maintained at the University of Helsinki \citep[\url{http://ipshocks.fi;}][]{kilpua2015properties}. SIR identifications are cross-checked against the ACE SIR catalog \citep[\url{https://izw1.caltech.edu/ACE/ASC/DATA/level3/SIR_List_1995_2009_Jian.pdf};][]{Jian2006}. From in situ data, we note that we identify HSSs as regions of $v_r \ge 500$~km~s$^{-1}$, or more generally as solar wind structures propagating faster than the given ICME at 1~au \citep{cranmer2017}. 

When possible, we complement in situ observations with WSA-ENLIL simulations of the ambient solar wind available on the NASA CCMC server (\url{https://ccmc.gsfc.nasa.gov}) to verify the presence of other solar wind structures that might have interacted with the ICME prior to its arrival to 1~au. 
It is important to keep in mind that in WSA-ENLIL simulations, the modeled arrival times of HSSs and the HCS at Earth location are not always consistent with the arrival times observed by spacecraft in situ \citep[see, for example,][]{gressl2014}. In this study, we therefore use WSA-ENLIL results as a guide on the global structure of the ambient wind, working under the assumption that modeled HSS and HCS arrival times are approximately correct. Further to the identification criteria listed in Section \ref{sec:identification}, we note that the magnetograms required for ENLIL runs are only available for one event in our study.

\section{Identification of Events} \label{sec:identification}

We start our selection process from the HELIO4CAST ICME catalog \citep[ICMECATv2.0;][]{moestl2017modeling,moestl2020prediction}, which incorporates the Wind ICME catalog of \citet{nieves2018understanding} with new entries identified by C. M\"ostl (a total of 436 ICMEs).  
\citet{belov2015galactic} found that a local minimum in the measured GCR flux (i.e. a Fd) was observed in association with ICMEs with peak magnetic field strengths typically $>$18~nT ($>$20~nT for ICMEs with MC signatures). Based on these findings, here we only consider MEs with clear FR signatures (i.e. compatible with the MC events considered by \citet{belov2015galactic}), and with a mean magnetic field strength greater than 20~nT (only 17 of the initial 436 ICMEs).
For the purpose of this work, we further narrow down the selection to four ICME events, selected according to the following criteria.
Two events are ``typical'' (or ``average'') ICMEs in terms of their in situ propagation and expansion speed, while two are ``fast'' ICME events, i.e. presenting outstanding propagation and expansion speeds. 
As a reference for the selection of our average ICME events, we note that \citet{richardson2010near} report average (mode) ICME transit and expansion speeds of 
$480$~km~s$^{-1}$ ($\sim 400$~km~s$^{-1}$) and $30$~km~s$^{-1}$ ($\sim 10$~km~s$^{-1}$) at 1~au, respectively.

Additionally, as we want to explore the relationship between interplanetary reconnection and Fd signatures, for each class of events (i.e. average vs fast) we select one event that presents predominantly BDE signatures throughout the FR (BDE fraction larger than 70\%), and one event that presents little or no BDE signatures (BDE fraction smaller than 5\%). For the identification of BDE vs non-BDE signatures, we employ the methodology presented in Section~\ref{subsec:BDE_identification}, and apply it to both ACE and Wind data for all the identified events, in order to have an idea about the uncertainty related to the use of a specific instrument/detector. To ensure the identification using the automatic detection algorithm is consistent with visual expectations and is not dependent on the specific visualization considered, we conduct a visual inspection of the suprathermal pitch angle distribution data plotted in two forms: one where the particle fluxes are colored using a logarithmic color scale, and one where the fluxes are normalized over time between a minimum value of 0 and maximum value of 1, and colored using a linear color scale.

We note that no identification conditions are applied to the associated Fd itself when selecting the ICMEs for this study, only that a decrease in GCR count from the background level associated with the passage of the ICME occurs. 

\section{Description of the Selected Events} \label{sec:description}

In the following sections, we present the four ICME events selected in this study (divided into pairs of ``average'' ICME speed and ``fast'' ICME speed), including descriptions of their solar eruption counterparts, interplanetary propagation, and plasma, magnetic field, and GCR signatures near Earth. We start with a discussion of the two average ICME events (Sections~\ref{subsec:average_event1} and \ref{subsec:average_event2}), and then move to the description of the two fast ICME events (Sections~\ref{subsec:fast_event1} and \ref{subsec:fast_event2}).

\subsection{Average Event 1} \label{subsec:average_event1}

\begin{figure*}
\centering
{\includegraphics[width=\textwidth]{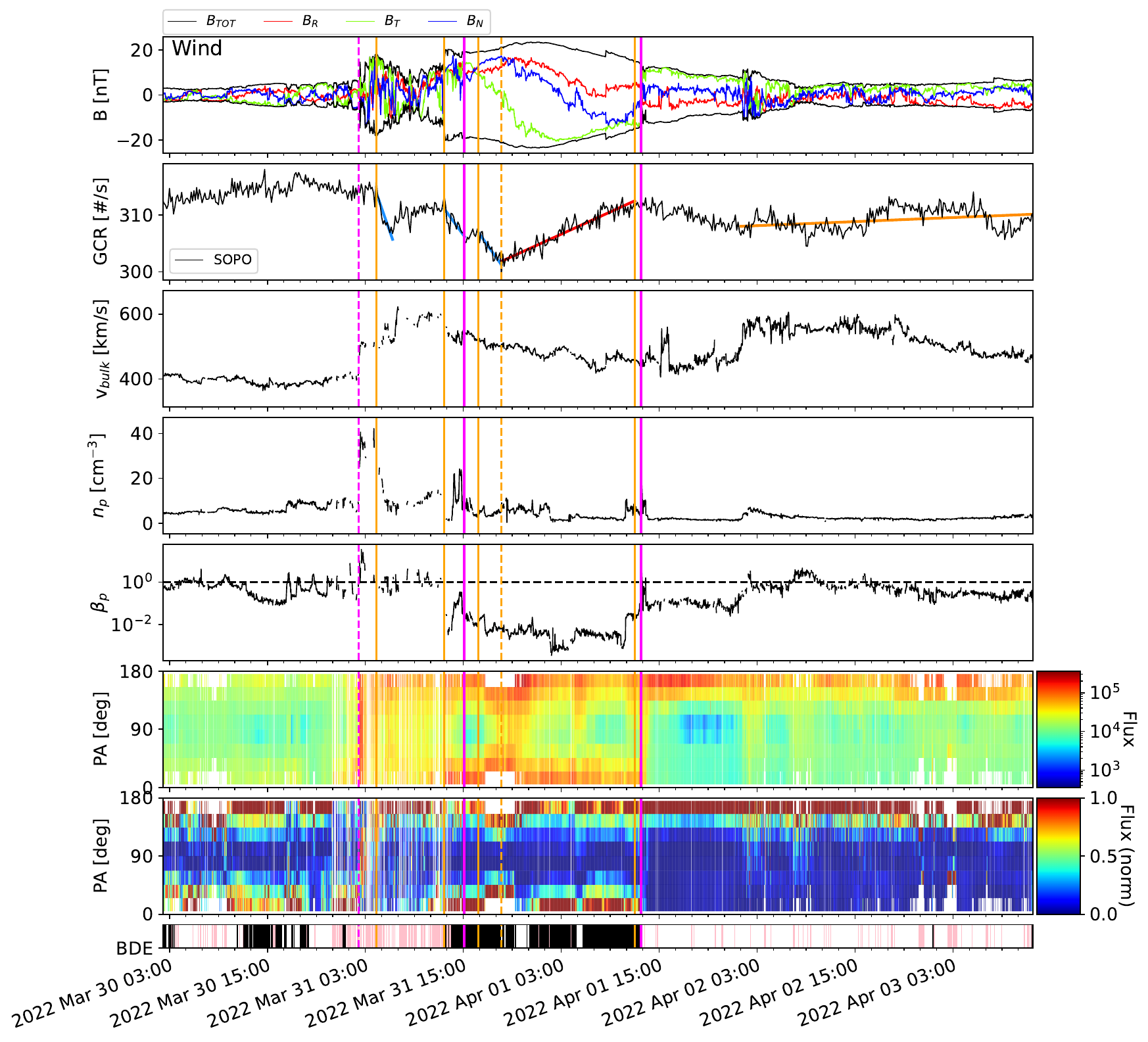}}
\caption{
In situ data for the ICME on 2022 March 31 -- April 1.
The first panel shows the Wind magnetic field data in RTN coordinates, with $B_R$ in red, $B_T$ in green, and $B_N$ in blue. The magnetic field magnitude $B$ and its negative $-B$ are in black. The same color coding for the magnetic field data is used in the following figures as well.
The second panel shows the SOPO neutron monitor data. The linear fits to the Fd steps, initial recovery and second recovery are presented by the light blue, red, and dark orange lines, respectively. 
The third, fourth, and fifth panels show the Wind proton bulk speed, number density, and proton $\beta$, respectively. 
The sixth and seventh panels show the Wind suprathermal electron PAD between 136~eV and 2~keV in normal and normalized forms (normalizing the maximum intensity at each time stamp to 1). 
The eighth plot reports the periods of BDE detection (in black), while data gaps in PAD are indicated in pink. 
The vertical magenta lines denote the shock time (dashed line) and the FR (solid lines) boundaries, while the orange vertical lines denote the SOPO Fd start, step times, and the end of the initial recovery associated with the passage of the FR (solid), respectively, and the minimum of the Fd (dashed). 
Wind data and ICME boundaries have been time-shifted to Earth as described in Section~\ref{subsec:time_shifting}.
} 
\label{fig:ae1}
\end{figure*}

The first event (hereafter AE1) is an ICME with average speed observed at Wind on 2022 March 31 -- April 1 and whose FR signatures are dominated by the presence of BDEs.
This ICME originated from active region NOAA AR 12975 on 2022 March 28. There were two CME eruptions from this region in fairly close succession on this day (the first CME was observed by LASCO C2 at 12:00 UT and the second observed by LASCO C2 at 20:24 UT) with the second CME having been $\sim$200 km~s$^{-1}$ faster than the first. As such, it is likely that the two CMEs interacted in interplanetary space well before reaching 1~au, as also confirmed by STEREO-A coronagraphic and heliospheric images and ENLIL simulations from the Space Weather Database Of Notifications, Knowledge, Information (DONKI) catalog (\url{https://kauai.ccmc.gsfc.nasa.gov/DONKI/}). Furthermore, ENLIL simulations also suggest that during its propagation from the Sun to the Earth, the ICME likely interacted with a concurrent HSS and its preceding HCS. Such an HSS is observed at STA on March 30--31 and the same HSS is later observed at Earth and visible in Figure~\ref{fig:ae1} at the back of the ME on April 2. Overall, both the observations and ENLIL simulations suggest that the interaction between the ICME and the HSS primarily involved the ICME portion east of the Sun--Earth line. It is intriguing that despite this interaction, and despite the fact that this ICME is associated to two interacting CMEs near the Sun, the FR remained topologically connected to the Sun through both legs, as evidenced by the extended BDEs shown in Figure~\ref{fig:ae1}. As such, we expect its ability to shield GCRs to be largely unaffected by these interactions.

Figure~\ref{fig:ae1} presents the time-shifted in situ data observed at Wind between 2022 March 30 -- April 2. Vertical magenta lines in the figure mark the arrival of the ICME forward shock, and the FR start and end, while Table \ref{tab:icme_properties} lists the corresponding times. Properties of the shock, sheath, and FR for this event are also given in Table \ref{tab:icme_properties}. We note that as this event occurred after 2018, the associated shock is not listed in the IPshocks catalog, and we thus extracted the shock properties following the documentation provided in the IPshocks database. The sheath mean speed of 548~km~s$^{-1}$ and the FR mean speed of 481~km~s$^{-1}$ make this an average speed ICME with a peak magnetic field strength of 23.6~nT. Based on the FR duration and mean speed, we estimate that the FR has a radial size of $\sim$0.25~au at 1~au. Within the FR, $B_N$ rotates from negative to positive and $B_T$ stays mostly negative in the front region of the FR. These rotations suggest this is a right-handed flux rope with an intermediate inclination between a WNE and a NES type \citep{bothmer1998structure}. Suprathermal electron PAD data indicate that there are BDEs for $\sim$75\% of the FR duration.

\input{icme_properties}
 
The Fd associated with this ICME at Earth is a three-step Fd as marked by the first three solid orange vertical lines in Figure~\ref{fig:ae1}. The Fd begins on 2022 March 31 at 04:20 UT, shortly after the shock front is observed. Following this initial decrease, there is a short recovery period during the sheath before a second decrease is observed almost coincident with the sudden increase in magnetic field strength (the start of the ME, rather than the FR delineated by the solid magenta vertical line). This decrease continues until a very small recovery in the GCR count rate is observed, around the time the smooth rotation of magnetic FR inside the ME begins. A third and final decrease is then observed on March 31 at 16:50 UT, until the Fd minimum at 19:40 UT, delineated by the dashed orange vertical line. The total percentage decrease in GCR count rate between the start of the Fd and the minimum is 5.4\%. We linearly fit each GCR step decrease within the Fd (shown by the light blue lines overlaying the GCR data in the second panel of Figure \ref{fig:ae1}), and the two rates of recovery observed (e.g. the recovery within the ME, shown in red, and the longer recovery toward the background GCR count rate after the passage of the ICME, shown in dark orange). The gradient of each slope fitted, and the percentage decrease in GCR count rate attributed to each step of the Fd, are given in Table \ref{tab:fd_properties}.

\input{fd_properties}

We note that the Fd presents an asymmetric profile one would expect due to ICME expansion, as shown by the steeper gradient of the third slope within the FR, in comparison to the gradient of the initial recovery. This finding is in line with the observation that the ICME is moderately expanding with an expansion speed of $\sim$52 km~s$^{-1}$. It is also important to note that the GCR recovery rate following the ICME has a shallower gradient than the initial recovery rate within the ME. This has previously been attributed to the ``shadow effect'' caused by the strength of the shock.
We further discuss possible relations between this long-lasting GCR flux recovery phase and the shock properties in relation to the ``shadow effect'' for this and the following events in Section~\ref{sec:discussion}. 

\subsection{Average Event 2} \label{subsec:average_event2}

\begin{figure*}
\centering
{\includegraphics[width=\textwidth]{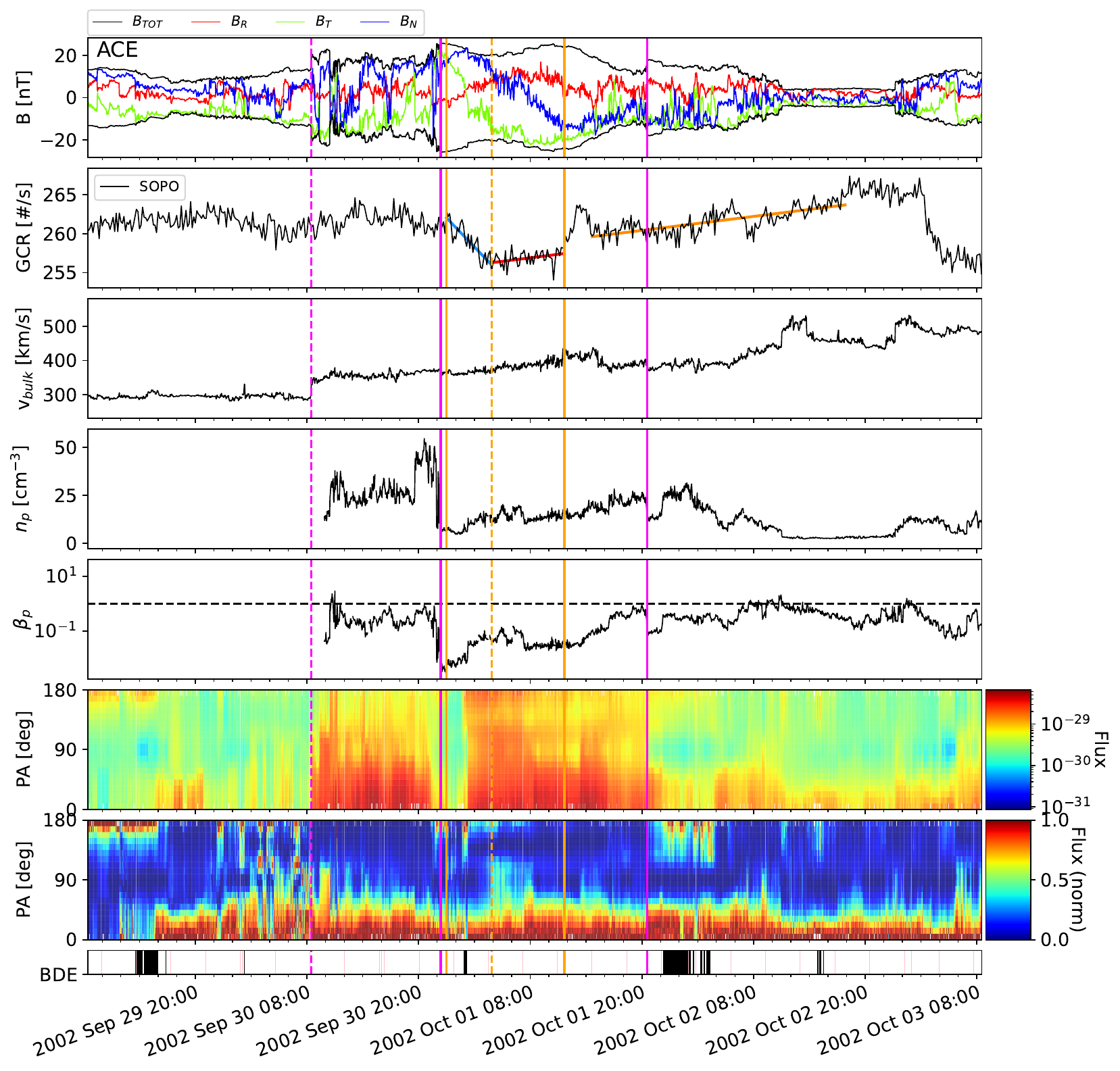}}
\caption{
In situ data for the ICME on 2002 September 30 -- October 1.
The first panel shows the ACE magnetic field data in RTN coordinates, using the same colors as in Figure~\ref{fig:ae1}. The second panel presents the SOPO neutron monitor data and fits to the Fd steps and recoveries, as in Figure \ref{fig:ae1}. The third, fourth, fifth, and sixth panels show the ACE proton bulk speed, number density, temperature, and proton $\beta$, respectively. The seventh, eighth, and ninth panels show the ACE suprathermal electron pitch angle distribution at 272 eV and the BDE detection in a similar format as in Figure~\ref{fig:ae1}. The vertical lines denote the shock, FR, and SOPO Fd step, minimum and FR recovery end times as in Figure~\ref{fig:ae1}. ACE data and ICME boundaries have been time-shifted to Earth as described in Section~\ref{subsec:time_shifting}.
}
\label{fig:ae2}
\end{figure*}

The second event (hereafter AE2) is an ICME with average speed observed at ACE on 2002 September 30 -- October 1, and whose FR signatures are dominated by the absence of BDEs.
Given the ICME impact speed around 387~km~s$^{-1}$ we estimate it would have taken about 4 days to propagate from the Sun to 1~au, so that its eruption is likely to have taken place on September 26 or early September 27. Several CMEs were reported within this time period, but they either lacked clear front-sided low coronal signatures (making it difficult to establish whether they were actually Earth-directed), or they were launched from source regions close to the limb (which do not easily justify an impact of the resulting ICME at Earth). Overall, univocal identification of the source region and CME counterpart for this ICME event is an extremely challenging endeavor due to the lack of multi-spacecraft remote-sensing observations, and the high level of activity of the Sun. 
Unfortunately, no ENLIL simulations were performed for the period associated with this ICME, likely due to the event occurring at a time when no high-resolution magnetograms needed to initialize ENLIL simulations were available (e.g. GONG or HMI). This implies that for the identification of possible interactions with other large-scale structures, we are limited to considering the in situ observations at ACE. In situ, we can see from Figure~\ref{fig:ae2} the presence of a faster structure compressing the ME from its back. We identify this structure as a HSS that could have originated from a coronal hole visible on the solar disk around $20^\circ$ south of the solar equator and passing through the central meridian on September 7. The SIR driven by this HSS is also listed in the ACE SIR catalog as embedding the ICME (starting on September 30 around 07:00 UT and ending on October 2 around 12:00 UT). No evidence of HCS crossings are present in the days preceding and following the ICME passage.
The multiple interactions underwent by this ICME during propagation from the Sun to 1~au provide a likely explanation for the mostly open magnetic topology exhibited by the ICME at ACE (lack of BDEs, and predominant presence of a single strahl, as shown in Figure~\ref{fig:ae2}).

Figure~\ref{fig:ae2} presents the time-shifted in situ data observed at ACE between 2002 September 29 -- October 2. Vertical magenta lines in the figure mark the arrival of the ICME forward shock, and the FR start and end. Similarly to AE1, these times are listed in Table \ref{tab:icme_properties} in addition to the properties of the shock, sheath, and FR. We note that due to the temporary data gap in the plasma properties, the interplanetary shock at ACE is not listed in the IPshocks catalog. Therefore, we consider information about the shock properties at Wind, which at that time was located about $8.7 \times 10^8$~m (about 136 Earth radii) away from ACE, and we list the properties in Table \ref{tab:icme_properties}. The sheath mean speed of 360~km~s$^{-1}$ and the FR mean speed of 387~km~s$^{-1}$ make this an average speed ICME with a peak magnetic field strength of 25.7~nT. Based on the FR duration and mean speed, we estimate that the FR has a radial size of $\sim$0.21~au at 1~au. 
Within the FR, $B_N$ rotates from north to south, and $B_T$ points mostly towards the east although a rotation from west to east is visible close to the FR front. These rotations suggest this is a right-handed low-inclination flux rope of NES type. The moderate $B_R$ intensity indicates the ICME is likely crossed near its nose and at a small impact parameter. 
The FR clearly lacks BDEs (present for only 1\% of the FR), while it presents electron strahl signatures predominantly around a pitch angle of $0^\circ$.

The second panel of Figure~\ref{fig:ae2} presents the neutron monitor data measured by SOPO during the event. No decrease in GCR count rate is observed around the arrival time of the ICME shock or during the sheath. The Fd is observed to start on 2002 September 30 at 23:00 UT and can be associated with the start of the FR. The Fd minimum is observed approximately 4~hours later, on October 1 at 03:50 UT. The total percentage decrease is 3.1\%. 

The recovery inside the FR is small, with a low gradient of $4.18 \times 10^{-5}$~counts~s$^{-2}$, listed in Table \ref{tab:fd_properties}. Most of the recovery in GCR count rate occurs towards the back of the FR, almost coincident with a sudden increase in velocity before the end of the smooth rotation of the magnetic field components. We observe a similar rate of recovery inside the FR to that outside following the passage of the FR; a very low positive gradient of $4.15 \times 10^{-5}$~counts~s$^{-2}$. The lack of recovery inside the FR in comparison to AE1, and thus the difference in shape of the Fd profile i.e. not asymmetric like AE1, may be due to the negative expansion rate (i.e. compression) within the FR of about $-10$~km~s$^{-1}$.

\subsection{Fast Event 1} \label{subsec:fast_event1}

\begin{figure*}
\centering
{\includegraphics[width=\textwidth]{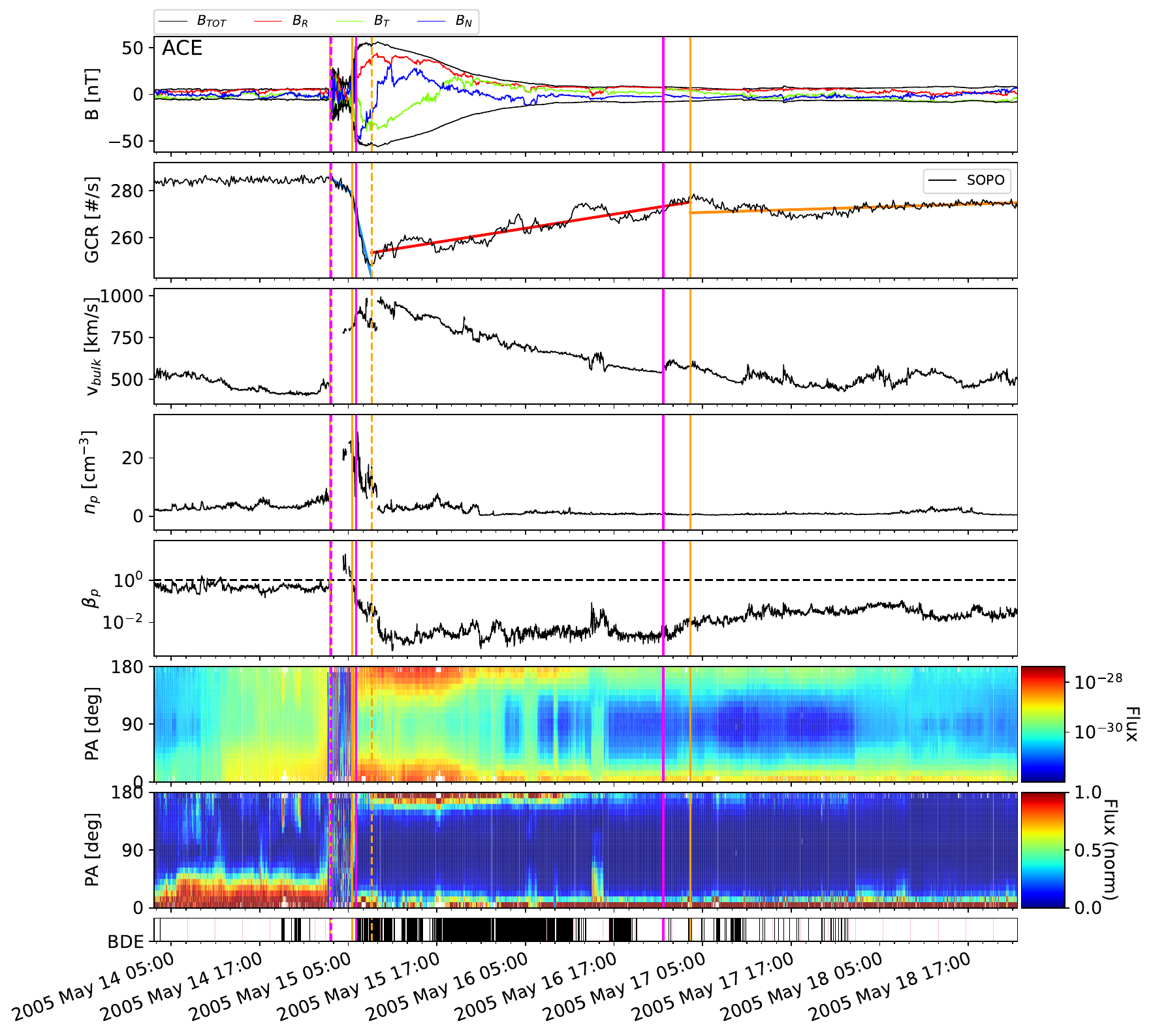}}
\caption{
In situ data for the ICME on 2005 May 15--17. 
ACE data and SOPO neutron monitor data are shown in the same format as in Figure~\ref{fig:ae2}. 
ACE data and ICME boundaries have been time-shifted to Earth as described in Section~\ref{subsec:time_shifting}.
}
\label{fig:fe1}
\end{figure*}

The third event (hereafter FE1) is a fast ICME observed at ACE on 2005 May 15 -- May 16 and whose FR signatures are dominated by the presence of BDEs.
This ICME was associated with a halo CME erupted from NOAA AR 10759 on 2005 May 13 (first observed in LASCO C3 at 17:12 UT). Similarly to AE2, due to its occurrence in Solar Cycle 23 when no high-resolution magnetograms were available, we cannot make use of reliable ENLIL simulations to reconstruct the propagation history of this ICME. Visual inspection of the in situ time series at ACE reveals a HCS crossing on May 13 around 16:30 UT (not shown), as indicated by the change of polarity in the solar wind $B_R$ component from positive to negative, accompanied by the change in the electron strahl direction from $180^\circ$ to $0^\circ$. The HCS crossing is, as expected, also accompanied by an increase in the proton density and $\beta$. As this takes place more than 1 day prior to the ICME arrival at ACE, however, we consider there was likely no interaction of the ICME with the HCS, at least in the vicinity of the Earth. In situ data also present no evidence that the ICME interacted with any SIRs/HSSs during its interplanetary propagation to 1~au. This scenario where the ICME propagated mostly undisturbed through interplanetary space, is consistent with the prolonged BDE signatures visible in Figure~\ref{fig:fe1} and a likely indicator of closed FR magnetic topology, i.e., the ICME being connected to the Sun via both legs.

Figure~\ref{fig:fe1} presents the time-shifted in situ data observed at ACE between 2005 May 15 -- May 17 in the same format as Figure~\ref{fig:ae2}. Similarly, all times and ICME properties are listed in Table \ref{tab:icme_properties}. We note that as for the previous case, a temporary data gap in the plasma properties prevents the characterization of the interplanetary shock properties at ACE. Therefore, we consider information about the shock properties at Wind, which at that time was located about $5.9 \times 10^8$~m (about 92 Earth radii) from ACE, and we list the properties in Table \ref{tab:icme_properties}. The sheath mean speed of 804~km~s$^{-1}$ and the FR mean speed of 777~km~s$^{-1}$ make this a fast ICME with a considerable peak magnetic field strength of 56.2~nT. Based on the FR duration and mean speed, we estimate that the FR has a radial size of $\sim$0.59~au at 1~au. Such an estimate falls among the largest reported \citep{richardson2010near}, and due to the outstanding expansion speed (around 152 km~s$^{-1}$), it could be a significant underestimation compared to reality. 
Within the FR, $B_N$ rotates from negative to positive and $B_T$ is mostly negative (eastward) in the front region of the FR but rotates to positive values (westward) in the second half. Rotations suggest this was a left-handed low-inclination flux rope of SEN type.
The intense $B_R$ indicates that the spacecraft crossed the ICME off-center or near its flank. About 73\% of the FR presents clear BDE signatures. 

The GCR count rate measured at SOPO is presented in the second panel of Figure \ref{fig:fe1}. We observe a steady initial decrease from 2005 May 15 at 02:30 UT, which is coincident with the start of the ICME, before a large main decrease on May 15 at 05:30 UT. The minimum of the Fd at SOPO occurs on May 15 at 08:10 UT, and we calculate a total percentage decrease of 13.6\% between the Fd start and minimum. The profile of the Fd is very asymmetric, similarly to the fast expanding profile of the FR. The time at which the maximum magnetic field strength within the ME is measured, on May 15 at 08:58 UT, occurs 48 minutes later than the time at which the Fd minimum is observed.  

Similarly to Sections \ref{subsec:average_event1} and \ref{subsec:average_event2}, the Fd slopes have been linearly fitted, as well as the recovery within the FR and after the passage of the FR. The recovery within the FR is variable and lasts until approximately May 17 at 03:20 UT. To reach the original background level, the following recovery rate is slower and takes until approximately May 22 at 15:30 UT (not shown in Figure \ref{fig:fe1}). At the transition between recovery rates, no obvious features in the magnetic field data or electron strahl are coincident; the electron strahls remain bi-directional at this time. However, the large recovery in GCR count rate between May 16 at 10:40 UT and 17:20 UT is almost coincident with a drop in intensity of electron strahl at a pitch angle of 180\dgg.

\subsection{Fast Event 2} \label{subsec:fast_event2}

\begin{figure*}
\centering
{\includegraphics[width=\textwidth]{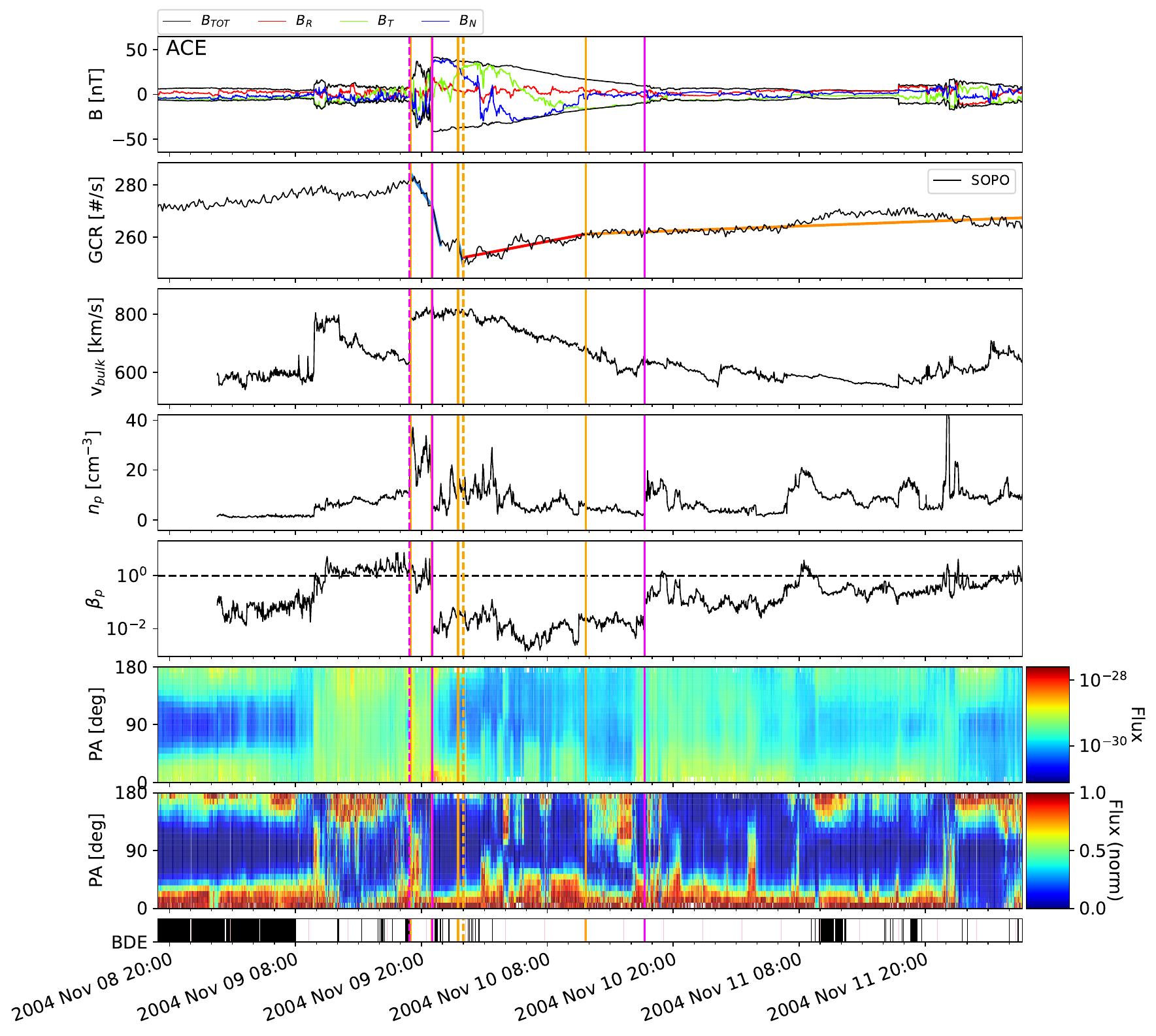}}
\caption{
In situ data for the ICME on 2004 November 9--10. 
ACE data and SOPO neutron monitor data are shown in the same format as in Figure~\ref{fig:ae2}.
ACE data and ICME boundaries have been time-shifted to Earth as described in Section~\ref{subsec:time_shifting}.
}
\label{fig:fe2}
\end{figure*}

The last event (hereafter FE2) is a fast ICME observed at ACE on 2004 November 9 -- November 10 and whose FR signatures are dominated by the lack of BDEs.
This ICME was associated with a halo CME erupted from NOAA AR 10696 on 2004 November 7 (first observed in LASCO C2 at 16:54 UT). Similarly to AE2 and FE1, due to its occurrence in Solar Cycle 23 when no high-resolution magnetograms were available, we cannot make use of reliable ENLIL simulations to reconstruct the propagation history of this ICME. 
Visual inspection of the in situ time series at ACE reveals the passage of an interplanetary shock preceding the ICME (detected on 2004 November 9 at 09:14 UT). The solar wind conditions after this first shock and until the second shock (driven by the ICME under investigation) have characteristics compatible with that of a sheath. This preceding shock is most likely caused by the passage of a preceding halo CME launched from the same AR on November 6, and observed as having a projected linear speed about 900 km~s$^{-1}$ slower than FE2 (as taken from the LASCO CME catalog). However, in situ there are no clear signatures of the passage of an ME after this first shock. Based on in situ observations, there is also no evidence that the ICME interacted with SIRs/HSSs or with the HCS during its propagation from the Sun to the Earth. 
It is possible that the interaction with the sheath or ejecta of a preceding ICME might have contributed to the magnetic disconnection of the FR, as suggested by the lack of BDE signatures and the detection of the electron strahl from alternating directions throughout the FR (Figure~\ref{fig:fe2}).

Figure~\ref{fig:fe2} presents the time-shifted in situ data observed at ACE between 2004 November 8 -- November 11, in the same format as the previous figures. Times and ICME properties are listed in the final column of Table \ref{tab:icme_properties}. The sheath mean speed of 788~km~s$^{-1}$ and the FR mean speed of 722~km~s$^{-1}$ make this a fast ICME with a considerable peak magnetic field strength of 41.7~nT. Based on the FR duration and mean speed, we estimate that the FR has a radial size of $\sim$0.35~au at 1~au (slightly higher than the average values reported in \citet{richardson2010near} and references therein). 
Within the FR, the $B_N$ magnetic field component rotates from positive to negative and $B_T$ rotates from positive (westward) to negative (eastward) values. These rotations are compatible with the crossing of a left-handed flux rope with intermediate inclination between a NWS and a WSE type. The very low $B_R$ magnitude indicates that the spacecraft likely crossed the flux rope near the center and near its central axis. 
The FR lacks consistent BDEs (present for only 3\% of the FR), and it contains periods of alternated strahl directions.

The SOPO data presented in the second panel of Figure \ref{fig:fe2} shows that the ICME drove a three-step Fd. The Fd is observed to start on 2004 November 9 at 19:00 UT, coincident with the ICME shock. The second Fd slope is coincident with the start of the FR, on November 9 at 21:00 UT. A third and final slope occurs within the FR on November 9 at 23:30 UT, following a slight recovery in the GCR count rate that occurs at a similar time to the start of the smooth rotation of the magnetic field components, especially B$_T$. The total Fd percentage decrease for the event is 12.1\%. We note that here we take the GCR count rate at the start of the Fd when calculating the total decrease, and not the background level, due to the preceding transient already affecting the GCR background.

The different recovery rates during and following the FR are shown by the red and dark orange lines overlaying the SOPO data. The recovery inside the FR is variable, continuing until November 10 at 11:40 UT around which time the smooth rotation of the magnetic field components within the FR ends. The following recovery continues at a slower rate and is fairly steady overall despite the following transients, finally recovering to the GCR background level on November 19 at 16:20 UT (taken as the background level from before the preceding transient to this event arrived, not shown in this figure). 

\section{Comparison of Events and Discussion} \label{sec:discussion}

As previously discussed, the mechanisms by which the different ICME substructures affect the modulation of GCRs can be differentiated: i) the ICME shock which reflects particles upstream causing an occasionally observed increase of GCR flux just prior to ICME arrival, and contributes to the ``shadow effect'' affecting the GCR recovery after the ICME has passed, ii) the diffusive barrier of the sheath affecting GCR transport which contributes to the initial decrease in GCR flux, and iii) the FR of the ICME often modeled with a closed magnetic field line geometry proposed to act as a shield to GCR transport contributing to the main decrease in GCR flux.

\subsection{Interaction history}

First, we investigate whether the modulation of GCRs is affected by the propagation history of the ICME, i.e., by the ICME having interacted with other transients during propagation. We address this aspect separately from the question of whether the reconnection of the ICME flux rope with other transients weakens its modulation of GCRs, as the interaction between ICMEs and other transients may not always involve extensive reconnection. We delve into this second question in the following section.

When considering the interaction with other large-scale structures in the interplanetary space, we found that three out of four events (AE2, FE1, and FE2) show a consistent pattern between their interaction history and the BDE signatures within the FR at 1 au. Specifically, one event which presented clear BDEs throughout the FR at 1 au (FE1) also propagated undisturbed from the Sun to the Earth, while the two events that exhibited a lack of BDEs within the FR at 1 au (AE2, FE2) also underwent significant interactions with other structures (and therefore possible extensive interplanetary reconnection). One event, however, did not fit this simple story: despite the likely interaction with a preceding CME and a following HSS, the FR of AE1 likely remained topologically connected to the Sun through both legs, as evidenced by the presence of extended BDEs. Based on these results, we argue that although some correlation between the interaction history of individual ICMEs and their magnetic connectivity was observed among the  small set of events considered, such a correlation is likely manifesting in a complex manner which makes it difficult to directly relate the interaction history to the expected ability of a given FR to shield GCRs.

\subsection{Effect of open/closed magnetic field topology on Fd properties}

Here, we test the hypothesis of the closed field line geometry mechanism, and thereby the effect of closed/open field lines on the FR's ability to modulate GCRs. We build on the results of \citet{richardson2011galactic}, where they found only a slight trend towards more magnetically closed ICMEs being associated with greater GCR decreases within the Fd; however, in their study, the fraction of closed field lines and the measured ICME speed were strongly correlated, thus making it difficult to determine which effect plays a more important role in GCR modulation within the FR. In this study, we selected four events with similar mean magnetic field strengths within the FR, but strong enough to drive a Fd \citep[$>$20~nT,][]{belov2015galactic}. The events in this study are split into two pairs: one pair of ICMEs with faster mean speeds around 750~km~s$^{-1}$, and one pair with more typical propagation speeds of 481 and 387~km~s$^{-1}$, respectively. Within each pair of ICMEs, we select one with a greater percentage of BDEs observed within the FR ($>$70\%), and one with mostly open field lines ($<$5\% BDEs). Such ICME properties for each event are summarized in Table~\ref{tab:icme_properties}. 

Comparing the first pair of events, ``Average Event 1'' (AE1) with a high percentage of BDEs and ``Average Event 2'' (AE2) with almost no BDEs, we initially observe very different Fd profiles. Three steps within the Fd can be identified at AE1 (marked in blue in Figure~\ref{fig:ae1}), the first corresponding to the start of the ICME, the second with the increased magnetic field strength and start of the BDEs (likely start of ME), and the third after the defined start of the smooth rotation of the FR. For each step, there is a similar decrease in GCR flux (3.12\%, 2.67\%, 2.72\%, respectively). However, the latter two steps decrease at half the rate of the first step at the start of the ICME sheath region ($-0.0011$, $-0.0005$, $-0.0005$~~counts~s$^{-2}$, respectively). The cumulative GCR flux decrease for these three steps is 5.44\% in comparison to the background level preceding the event, with the third step (i.e. the one within the defined FR) contributing to a decrease of 2.72\% in GCR flux in comparison to the background level preceding the event. For AE2, only one step in the Fd profile is clearly observed, which is almost coincident with the start of the FR. This step contributes to a decrease in GCR flux of 3.10\%. Overall, despite the lack of BDEs within the FR of AE2, we find a similar percentage decrease in the GCR flux due to the FR to that found for AE1.

Comparing the fast events, ``Fast Event 1''(FE1) with a high percentage of BDEs and ``Fast Event 2'' (FE2) with almost no BDEs, we also observe different Fd profiles. FE1 possesses a two-step Fd profile, with the first step corresponding clearly to the start of the ICME (coincident with the ICME shock), and the second decrease at the initial sharp increase in magnetic field and BDEs, just before the start of the defined FR. The second decrease in GCR flux associated with the FR is 11.18\%. Three steps can be identified in the Fd profile of FE2, the first coincident with the ICME shock, the second coincident with the FR start, and the third approximately two hours later within the FR. The two steps within the FR combined contribute to a total decrease in GCR flux of 8.04\%.

Comparing the events with BDEs to those without within each speed category would suggest that the presence of BDEs, or lack thereof, within the FR does not play a large role driving the Fd. This is in agreement with \citet{richardson2011galactic} where the total percentage GCR flux decrease within MEs is similarly variable across different BDE percentages until those $>$90\% \citep[see Figure 11 of][]{richardson2011galactic}. It would instead seem from our different event speed pairings that the faster events drive larger Fds; this general trend was also observed in \citet{richardson2011galactic}. However, we note that they still reported a large variability in the total percentage GCR flux decrease detected within MEs of different speeds \citep[see Figure 9 of][]{richardson2011galactic}.

We also cannot disentangle the role the magnetic field strength plays within our event speed pairings in this study, as both fast events also have much greater peak magnetic field magnitudes. This is unsurprising, as the ICME maximum magnetic field strength is well-known to be correlated with the interplanetary speed of the ICME \citep[e.g.][]{moestl2014}.
As mentioned in the introduction, GCR diffusion models across the ME find the Fd size proportional to $Bva^2$; therefore, it is possible that the combination of the high speed and high peak magnetic field strength both played a significant role in causing FE1 and FE2 to have larger Fds than AE1 and AE2. We also note, however, that \citet{richardson2011galactic} found a similarly large variability in the total percentage GCR flux decrease within MEs of various magnetic field strengths. Interestingly, \citet{richardson2011galactic} found a slightly stronger correlation between ICME speed and total percentage GCR flux decrease ($-0.64$), than magnetic field strength and total percentage GCR flux decrease ($-0.48$), thus suggesting that speed may still play a larger role in driving the Fd. 

\subsection{Smaller-scale effects of open/closed magnetic field boundaries}

Moving on from general properties of ICMEs and Fds, we now focus on the smaller scale details of each event. 
We first investigate whether there are any periods of recovery in the GCR flux during short periods of open field lines within the FR \citep[as suggested in][]{bothmer1997effects}. This is possible to investigate for the events with mostly BDEs (AE1 and FE1), as there are brief periods within the FRs where BDEs are not indicated (see bottom panel of Figures \ref{fig:ae1} and \ref{fig:fe1}).

For AE1, Figure \ref{fig:ae1} indicates two periods without BDEs between 2022 March 31 18:00 -- 19:00 UT and April 01 04:00 -- 04:30, however, there is no observable affect or change to the GCR recovery within the SOPO data for this event. FE1 also presents two brief periods not indicated as BDEs, between 2005 May 15 11:00 -- 13:00 UT, and May 15 15:00 -- 17:00 UT. These occur in close succession to each other and we observe a recovery in the GCR flux during the first period, which is, however, also coincident with a brief bump in the speed, radial component of the magnetic field, and plasma beta. Following these features, the GCR flux returns to its previous level (decreases again) and continues to recover at the same rate as before during the second period where no BDEs are indicated, towards the end of the FR.

Recoveries of the GCR flux within the FR do not occur exclusively for the events with only brief periods of open field lines. For example, Figure \ref{fig:ae2} for AE2 shows a comparatively large recovery in GCR flux within the FR between 2022 October 1 12:00 and 15:00 UT. This recovery occurs when the field lines are considered completely open, but is also coincident with a rise and fall in the bulk velocity data. It is therefore not conclusive whether there is a close association between small recoveries in the GCR flux observed and brief periods of open field lines within the FR, or whether the recoveries are caused mostly by other plasma and magnetic field features such as fluctuations or increases in the FR speed itself. 

For events AE1 and FE1, we are also able to investigate whether there is an obvious change in GCR flux between the closed field lines within the FR and the open field lines outside of the FR. At the defined FR end boundary of AE1 (see Figure \ref{fig:ae1}), we observe a clear boundary between closed to open field lines. At this boundary, instead of a change in recovery rate of the GCR flux i.e from a faster recovery rate inside the FR to a slower recovery rate after the passage of the ME, we actually observe a decrease in GCR flux. This decrease occurs over the same time period in which there is an increase in the proton $\beta$ (although $\beta$ is still less than one during this period), and ends with the arrival of faster solar wind plasma which causes a further decrease in GCR flux. 
The behavior at the boundary of FE1 is more complicated (see Figure \ref{fig:fe1}), with a period of open field lines from 2005 May 16 11:00 to 16:00 UT, either side of the defined FR boundary. During this time, we do observe a bump in the GCR recovery rate, as well as an increase in the proton $\beta$ and bulk velocity, before returning to the previous recovery rate until May 17 03:20 UT. At this time, we observe again a decrease in GCR flux, coincident with the proton $\beta$ starting to rise, before a more consistent slower recovery rate begins on May 22 15:30 UT. 

\subsection{Comparison of recovery rates}

Two different recovery rates, one starting from the minimum of the GCR flux within the FR (and possibly continuing beyond the defined FR end boundary), and a second after the passage of the FR, are common across both fast events (FE1 and FE2) and one average event (AE1), although the two separate recovery periods can also be distinguished for AE2 despite recovering at similar rates due to the bump in GCR flux previously mentioned. The different recovery rates and duration are given in Table \ref{tab:fd_properties}. For FE1, FE2 and AE1, the initial recovery rate surrounding the passage of the FR is steeper than the second recovery rate outside of the FR. Comparing events within speed categories, for FE1 and FE2, even though both events have similar percentage minima reached in the GCR flux (13.6\% and 12.1\%, respectively), FE2 (mostly open field lines) recovers at a faster rate within the FR than FE1 (mostly closed field lines). This is the opposite case for AE1 and AE2, where the recovery rate within the FR of AE1 (mostly closed field lines) is much faster than that of AE2. Considering that two distinct recovery rates are observed regardless of the BDE percentage within the FR, this difference would therefore suggest that it is not the ending of the closed magnetic field lines of the FR that causes the difference in recovery rates. 

The second recovery rate outside of the FR is thought to be due to the ``shadow effect'', where the recovery would be affected by the shock front reflecting the upstream particles. From this, we would expect that given similar ICME properties (in terms of propagation speed and angular/radial sizes), a stronger shock would reflect more particles, and therefore, be associated with a slower GCR flux recovery phase. For the average events, we calculate a shock magnetosonic Mach number of 1.0 and 1.1 for AE1 and AE2, respectively (as reported in Table~\ref{tab:icme_properties}). Despite similar Mach numbers, we find the recovery of AE2 to be 2.5 times faster than the rate of AE1. One explanation for this difference may be related to the fact that AE2 is less extended in the radial direction than AE1 (total ICME radial size of about 0.33 au and 0.42 au, respectively; from Table~\ref{tab:icme_properties}). Assuming a similar ratio also hold in the angular directions, the shock driven by AE2 might be significantly less extended in the angular direction than the one of AE1, therefore contributing to a lesser shielding of the spacecraft from GCRs after the ME passage beyond the observer.
The fast events have very different Mach numbers, 4.7 for FE1 and 2.4 for FE2. For these events, the recovery rates follow the expected pattern: the recovery rate of FE2 is 1.6 times slower than that of FE1. Such a behavior is also consistent with the radial sizes of the ICMEs (from Table~\ref{tab:icme_properties}): FE1 has a total radial size of about 0.65 au, while FE2 has a total radial size of about 0.40~au. 
If a similar ratio in the angular sizes were to hold as well, it would further justify the faster recovery rate associated with FE2.

\subsection{Fd structure and profile}

Comparing the general Fd structures i.e. number of steps observed, we find no relationship between steps driven across the different speed pairings for events with or without BDEs. The events do, however, follow many of the mechanism expectations for the initial decreases, with AE1, FE1 and FE2 all displaying an initial decrease coincident with the ICME shock, and a second decrease almost coincident with the FR. The overall Fd profile within the FR is also very asymmetric for each of the events listed above. Previously, \citet{belov2015galactic} suggested that faster ICMEs produce more asymmetric Fds. In our study, we find that it is not just the fast events that produce asymmetric Fds, but also the more average speed event of AE1. The commonality between the events is the FR expansion speed: AE1, FE1 and FE2 have expansion speeds of 52, 152, and 88~km~s$^{-1}$, respectively. In an idealized model with a linearly decreasing speed profile within the FR, as shown by Figure 2 in \citet{dumbovic2020evolution}, an asymmetry is produced in the resulting Fd profile within the FR. The expansion speed of FE1 is almost twice that of FE2, and produces a GCR decrease in the FR with a gradient almost 6 times steeper than the initial decrease after the ICME shock. In comparison, the decrease rate within the FR is approximately 3 times steeper than that within the sheath for FE2. Interestingly, AE2 displays a very different Fd profile in comparison to AE1, FE1 and FE2. As described in Section \ref{subsec:average_event2}, the Fd profile is almost flat within the Fd after the initial decrease in GCR flux. We note that this is the only event to have a negative expansion speed within the FR (the speed profile linearly increases throughout the FR, likely compressed by the following high speed solar wind plasma), which likely contributes to the comparatively different Fd profile of this event.

\section{Conclusion} \label{sec:conclusion}

In this study, we have investigated the contribution of different proposed mechanisms to Fds driven by ICMEs at Earth, and attempted to discern whether closed magnetic field lines within the FR, or other features of the ICME such as speed and magnetic field strength, play a more important role in determining the size and structure of the Fd. To address these aims, we identified four ICMEs observed at L1 that drove Fds observed by neutron monitors at Earth: two pairs of events that differ in ICME speed i.e. one pair of ICMEs with more typical mean speeds and the other pair with faster mean speeds, and within each speed pairing, one ICME that exhibited an open magnetic field line topology and one that exhibited a mostly closed topology. We note that although the mean magnetic field strengths are similar across all events, the fast ICMEs have a much stronger maximum magnetic field strength than the more typical speed ICMEs, and thus, we cannot completely separate the effect the ME speed and peak magnetic field strength have on the resulting Fd.

It seems much of the GCR response to the ICME events in this study is independent of whether or not a high percentage of BDEs, i.e. closed magnetic field lines, are present within the FR. We therefore conclude that closed field lines within the FR of ICMEs plays less of a role in driving the Fd as previously thought, whereas features such as the fluctuations in FR speed and magnetic field components, and expansion play more of a role in determining the smaller-scale structure of the Fd profile. We hypothesize that perhaps such speed fluctuations may create compressions and rarefactions in the FR, or boundary layers, affecting the diffusion of GCRs. 

We also find that it is not only fast ICMEs that produce asymmetric Fds, but a positively expanding FR, in agreement with \citet{dumbovic2020evolution}. We suggest that perhaps the speed at which different parts of the ICME travel, and therefore, expansion, may also affect the rate at which GCRs can fill the magnetic structure of the FR, affecting both the larger scale and smaller scale structure of the Fd. 

Overall, our observational study suggests that although the mechanisms used in Fd models do a good job of describing overall GCR response, the real picture is not as simple. In the four events studied here, each ICME generated unique Fd that seemingly depended on small-scale features present in the ICME as well. More studies correlating smaller-scale magnetic and plasma features within MEs with the GCR response, for both events with a FR structure (like the events in this study), and those without a FR structure, are still needed to improve our understanding of the mechanisms that drive Fds. 

\begin{acknowledgments}

This research was supported by NASA grant 80NSSC19K0914 (E.E.D., C.S., and R.M.W.). 
C.S. also acknowledges NASA grants 80NSSC20K0197 and 80NSSC20K0700. R.M.W. was partially supported by the NASA STEREO grant 80NSSC20K0431. E.E.D and C.M. acknowledge funding by the European Union (ERC, HELIO4CAST, 101042188). Views and opinions expressed are however those of the author(s) only and do not necessarily reflect those of the European Union or the European Research Council Executive Agency. Neither the European Union nor the granting authority can be held responsible for them. \end{acknowledgments}

\bibliography{bibliography}{}
\bibliographystyle{aasjournal}

\end{document}

%% file: icme_properties.tex
\begin{table}
\centering
 \begin{tabular}{l|cc|cc} 
 \hline \hline
                                            & \textbf{Avg. Event 1} & \textbf{Avg. Event 2} & \textbf{Fast Event 1} & \textbf{Fast Event 2} \\
 \hline
{ICME start time (shock) [UT]}            & 2022-03-31 01:46$^\diamond$ & 2002-09-30 07:20$^\star$ & 2005-05-15 02:10$^\star$ & 2004-11-09 18:20$^\star$ \\
{FR start time [UT]}            & 2022-03-31 14:32$^\diamond$ & 2002-09-30 21:10$^\star$ & 2005-05-15 05:32$^\star$ & 2004-11-09 20:30$^\star$ \\
{FR end time [UT]}              & 2022-04-01 12:04$^\diamond$ & 2002-10-01 19:30$^\star$ & 2005-05-16 13:06$^\star$ & 2004-11-10 16:30$^\star$ \\
 \hline
 Shock speed [km/s]                         & 128$^\diamond$     & 345$^\diamond$     & 848$^\diamond$ & 816$^\star$  \\
 Shock speed jump [km/s]                    & 105$^\diamond$     & 42$^\diamond$      & 296$^\diamond$ & 147$^\star$  \\
 Shock $\theta_{Bn}$ [$^\circ$]             & 39$^\diamond$      & 72$^\diamond$      & 62$^\diamond$ & 14$^\star$   \\
 Shock magnetosonic Mach number             & 1.0$^\diamond$     & 1.1$^\diamond$     & 4.7$^\diamond$ & 2.4$^\star$  \\
 \hline
 Sheath mean (peak) speed [km/s]         & 548 (623)$^\diamond$    & 360 (382)$^\star$     & 804 (894)$^\star$ & 788 (832)$^\star$     \\
 Sheath mean (peak) magnetic field [nT]  & 12.4 (20.4)$^\diamond$  & 17.8 (25.7)$^\star$   & 18.5 (51.7)$^\star$ & 25.2 (41.0)$^\star$  \\
 Sheath duration [hrs]                      & 12.87$^\diamond$        & 13.83$^\star$         & 3.35$^\star$ & 2.15$^\star$          \\
 Sheath size [au]                           & 0.17$^\diamond$       & 0.12$^\star$        & 0.06$^\star$ & 0.05$^\star$            \\
 \hline
 FR mean (peak) speed [km/s]         & 481 (551)$^\diamond$     &  387 (442)$^\star$    & 777 (995)$^\star$ &  722 (853)$^\star$         \\ 
 FR expansion speed [km/s]              & 52$^\diamond$            &  -10$^\star$           & 152$^\star$ &  88$^\star$               \\ 
 FR mean (peak) magnetic field [nT]  & 20.1 (23.6)$^\diamond$   &  20.5 (25.8)$^\star$  & 27.3 (56.2)$^\star$ &  25.3 (41.7)$^\star$      \\
 FR duration [hrs]                      & 21.53$^\diamond$         &  22.33$^\star$        & 31.56$^\star$ &  20.0$^\star$             \\
 FR size [au]                           & 0.25$^\diamond$        &  0.21$^\star$       & 0.59$^\star$ &  0.35$^\star$           \\
 Fraction of FR with BDEs               & 75\%$^\diamond$          &  1\%$^\star$           & 73\%$^\star$ &  3\%$^\star$              \\
 \hline
 \end{tabular}
 \caption{Summary of ICME parameters. 
 $^\star$values at ACE;	
 $^\diamond$values at Wind.}
\label{tab:icme_properties}
\end{table}

%% file: fd_properties.tex
\begin{table}
\centering
\begin{tabular}{l|cc|cc} 
\hline \hline
                      & \textbf{Avg. Event 1} & \textbf{Avg. Event 2} & \textbf{Fast Event 1}  & \textbf{Fast Event 2}  \\ \hline
Fraction of ME with BDEs        & 75\%             & 1\%              & 73\%             & 3\%                \\ \hline
Fd Start {[}UT{]}               & 2022-03-31 04:20 & 2002-09-30 23:00 & 2005-05-15 02:30 & 2004-11-09 19:00 \\
1st Step Gradient {[}counts~s$^{-2}${]}  & $-1.15$E$-$03    & $-0.35$E$-$03     & $-0.61$E$-$03     & $-1.71$E$-$03     \\
1st Step Decrease {[}\%{]}          & 3.12      & 3.10      & 2.71            & 4.09            \\ \hline
2nd Step Start {[}UT{]}         & 2022-03-31 12:40 &  & 2005-05-15 05:30 & 2004-11-09 21:00  \\
2nd Step Gradient {[}counts~s$^{-2}${]} & $-0.55$E$-$03     &         & $-3.53$E$-$03     & $-5.18$E$-$03     \\
2nd Step Decrease {[}\%{]}          & 2.67      &                  & 11.18       & 5.44       \\ \hline
3rd Step Start {[}UT{]}         & 2022-03-31 16:50 & & & 2004-11-09 23:30 \\
3rd Step Decrease {[}counts~s$^{-2}${]} & $-0.56$E$-$03     &                  &                  & $-5.55$E$-$03     \\
3rd Step Decrease {[}\%{]}          & 2.72      &                  &                  & 3.80      \\ \hline
Total Decrease {[}\%{]}             & 5.44      & 3.10      & 13.58      & 12.13      \\ \hline
1st Recovery Gradient {[}counts~s$^{-2}${]} & 18.16E$-$05      & 4.18E$-$05         & 13.79E$-$05      & 21.42E$-$05      \\
1st Recovery Duration {[}hrs{]} & 16.3             & 7.8              & 43.2             & 11.7             \\
1st Recovery End {[}UT{]}       & 2022-04-01 12:00 & 2002-10-01 11:40 & 2005-05-17 03:20 & 2004-11-10 11:40 \\
2nd Recovery Gradient {[}counts~s$^{-2}${]} & 1.66E$-$05         & 4.15E$-$05         & 2.68E$-$05         & 4.16E$-$05         \\
2nd Recovery Duration {[}hrs{]} & 128              & 30.2             & 132.2            & 220.7      \\ 
2nd Recovery End {[}UT{]}       & 2022-04-06 20:00  & 2002-10-02 17:50 & 2005-05-22 15:30 & 2004-11-19 16:20 \\ \hline
\end{tabular}
\caption{Summary of Fd properties.}
\label{tab:fd_properties}
\end{table}